\documentclass[amssymb,aps,twocolumn, nofootinbib]{revtex4}
\usepackage[]{graphicx}
\usepackage[]{amsmath}
\usepackage{hyperref}
\usepackage{subfigure}

\def\II{1\!\mathrm{l}}

\def\cG{\mathcal{G}}

\def\cK{\mathcal{K}}

\def\cO{\mathcal{O}}
\def\cP{\mathcal{P}}

\def\cS{\mathcal{S}}
\def\cT{\mathcal{T}}

\def\eq#1{Eq.~\eqref{eq:#1}}
\def\fig#1{Fig.~\ref{fig:#1}}
\def\sec#1{Sec.~\ref{sec:#1}}
\def\be{\begin{equation}}
\def\ee{\end{equation}}
\def\BE{\begin{equation}}
\def\EE{\end{equation}}
\newcommand\BF{\begin{figure}[t!]}
\newcommand\EF[2]{\caption{#1}\label{#2}\end{figure}}

\begin{document}

\title{Fault-Tolerant Renormalization Group Decoder for Abelian Topological Codes}
\author{Guillaume Duclos-Cianci}
\affiliation{D\'epartement de Physique, Universit\'e de Sherbrooke, Sherbrooke, Qu\'ebec, J1K 2R1, Canada}
\author{David Poulin}
\affiliation{D\'epartement de Physique, Universit\'e de Sherbrooke, Sherbrooke, Qu\'ebec, J1K 2R1, Canada}

\date{\today}

\begin{abstract}
We present a three-dimensional generalization of a renormalization group decoding algorithm for topological codes with Abelian anyonic excitations that we introduced for two dimensions in \cite{DP09a,DP10a}. This 3D implementation extends our previous 2D algorithm by  incorporating a failure probability of the syndrome measurements, i.e., it enables fault-tolerant decoding. We report a fault-tolerant storage threshold of $\sim1.9(4)\%$ for Kitaev's toric code subject to a 3D bit-flip channel (i.e. including imperfect syndrome measurements). This number is to be compared with the $2.9\%$ value obtained via perfect matching \cite{H04a}. The 3D generalization inherits many properties of the 2D algorithm, including a complexity linear in the space-time volume of the memory, which can be parallelized to logarithmic time. \end{abstract}


\maketitle

\section{Introduction}

Topological quantum error-correcting codes currently stand as some of the most promising implementations of quantum memories and computers. Crudely, topological codes are standard quantum error-correcting codes  with additional geometric constraints: their check operators involve only neighbouring spins on a two dimensional (2D) lattice. As a consequence, they can exhibit high fault-tolerant threshold \cite{WFH10a, AKBM11a, LAR11a} with relatively low overhead. Some topological codes also support transversal implementation of Clifford gates \cite{BM06a}, which simplifies fault-tolerant quantum computation. Lastly, topological codes can be efficiently decoded \cite{DKLP02a,H04a,DP09a}, which is the topic of this paper.

Decoding a quantum code consists in inferring the optimal recovery given a statistical description of the noise  and an error syndrome---i.e., the measurement outcome of check operators which reveal incomplete information about the particular error that has affected the system. Thus, decoding is a classical statistical inference problem involving a very large number of correlated random variables. Extremely fast decoding algorithms are required to prevent errors from building up in between error correction cycles, although some lag-time can be tolerated, e.g., by extending ideas from \cite{AD09a}. In \cite{DP09a,DP10a}, we introduced a decoding algorithm for Kitaev's topological code \cite{K03a} that uses renormalization group (RG) techniques from statistical physics. It's complexity is linear with the number of qubits, as compared to the cubic complexity of previously known algorithms \cite{E65a}. Most importantly, it can be parallelized to logarithmic time. 

The present paper is a continuation of our work initiated in \cite{DP09a,DP10a}, and serves many purposes. 1) Our previous work focused on error correction in the presence of perfect syndrome measurements. When measurements are faulty, fault-tolerant techniques are required which change the nature of the decoding problem. As we explain below, for topological codes, this can be effectively described by increasing the lattice dimension by one dimension representing time \cite{DKLP02a}. Thus, we adapt our RG algorithm, initially devised for a 2D lattice, to a 3D fault-tolerant setting.\footnote{Note that we have used our algorithm in a fault-tolerant setting in \cite{BDPS12a}, but did not provide any details of the implementation.} 2) Our algorithm was devised specifically for Kitaev's topological code. Because all 2D stabilizer codes are locally equivalent to multiple copies of Kitaev's code \cite{BDP11a}, our RG algorithm can be used with any such code. However, this requires determining the local mapping that realizes this equivalence, and transforming the local noise model accordingly, which can in principle affect the decoder's performances. Here, we describe our methods in physical terms that are directly applicable to any code that supports Abelian anyons  \cite{B11a,BDP11a,B09a, K03a, DP13a}, not restricted to stabilizer codes. We have implemented a special case of this generalization in \cite{DP13a} for the $\mathbb Z_d$ quantum double model. 3) Our previous publications on this topic focused on applications, giving only a high level description of the actual algorithm. Here, we provide a complete detailed description of the structure of the algorithm, which should be sufficient for anyone interested in implementing it. 

The rest of the paper is organized as follows. In the next section, we provide a heuristic physical description of the algorithm in terms of localized Abelian anyons. This section should provide a good physical intuition of the different components of the algorithm. This is first done assuming perfect syndrome measurements, and in the last subsection we explain how the problem is modified in the presence of faulty errors, following \cite{DKLP02a}. Section~\ref{sec:KTC} revisits all the concepts introduced heuristically in \sec{heuristic} for the special case of Kitaev's topological code, using an algebraic formalism closely related to the actual implementation of the algorithm. Section~\ref{sec:numerics} presents our numerical experiments, and we conclude in \sec{conclusion} with possible extensions and relations to other methods. Appendix \ref{sec:marginal} details our mathematical notation for probability distributions over the $n$-qubit Pauli group.

\section{Heuristic physical description}
\label{sec:heuristic}

In this Section, we provide a heuristic physical description of the problem of interest, and of the numerical tools we have developed to solve it. A more detailed mathematical description is presented in \sec{KTC}.

\subsection{Decoding problem}

\begin{figure}
  \subfigure[ ]{
  \includegraphics[width=5 cm]{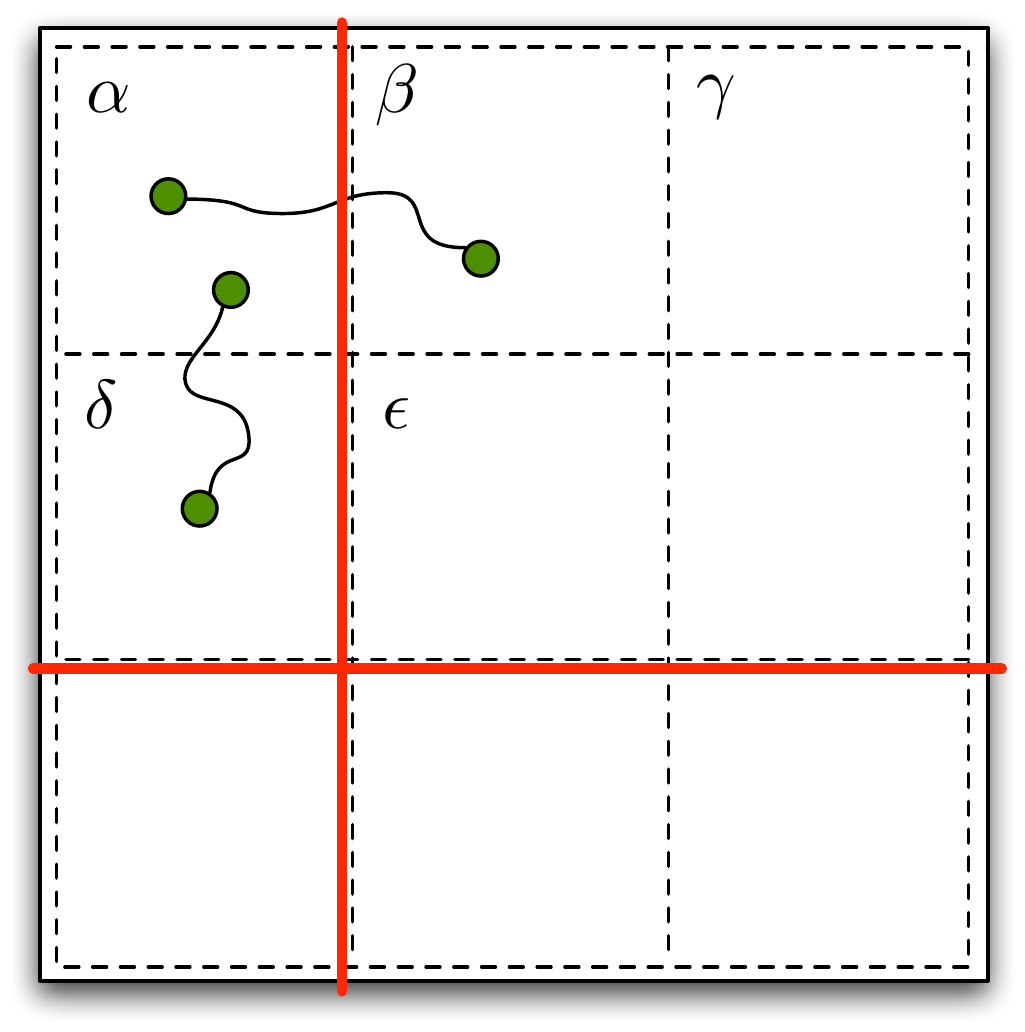}
  \label{fig:RGAL}}
  \subfigure[ ]{
  \includegraphics[width=2 cm]{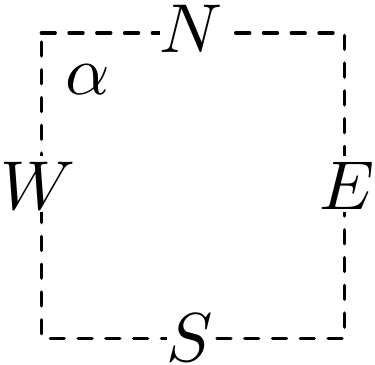}
  \label{fig:RGAR}
  }
  \caption{(a) A 2D topological code is cut into unit cells $\alpha$, $\beta$, ... Gauge lines representing the non-trivial cycles (solid red lines) are  chosen arbitrarily. Computing the flow of charge through the gauge lines is equivalent to decoding. (b) Each region has four boundaries that we label north ($N$), east ($E$), south ($S$) and west ($W$).}
  \label{fig:RGA}
\end{figure}

Consider a 2D sheet of topological matter supporting Abelian anyons. For simplicity, suppose that the system has periodic boundary conditions, so it forms a torus. The information is encoded in the degenerate ground state of the system. Excitations above the ground state manifold are localized Abelian anyons---they carry conserved charges $\{a,b,c , \ldots \}$ that obey ``deterministic" fusion rules, e.g. $a\times b = c$. The information in the ground state can be modified by creating a particle-antiparticle pair $(a, \bar a)$, dragging one of the particle around a topologically non-trivial cycle, and fusing it with its original partner $a\times \bar a = 1$. 

In the presence of errors, such a process could occur spontaneously. For instance, the creation of a particle-antiparticle pair could result from a thermal fluctuation. Once created, additional errors could cause the particles to diffuse on the sheet. To prevent  corruption of the memory, we must therefore keep track of the homology of the particles' world-lines. Periodic measurements of the particles' location yield partial information about their trajectories, and the {\em decoding problem} becomes one of statistical inference: it sets to determine the most likely homology of the particles' world-lines  given two consecutive snapshots of their locations. Concretely, we can arbitrarily choose two gauge lines representing the two non-trivial cycles of the torus [c.f. \fig{RGAL}], and the decoding problem consists in determining the net flow of charge, or current, across these two gauge lines. 

\subsection{RG algorithm}

\begin{figure}
\includegraphics[width=5cm]{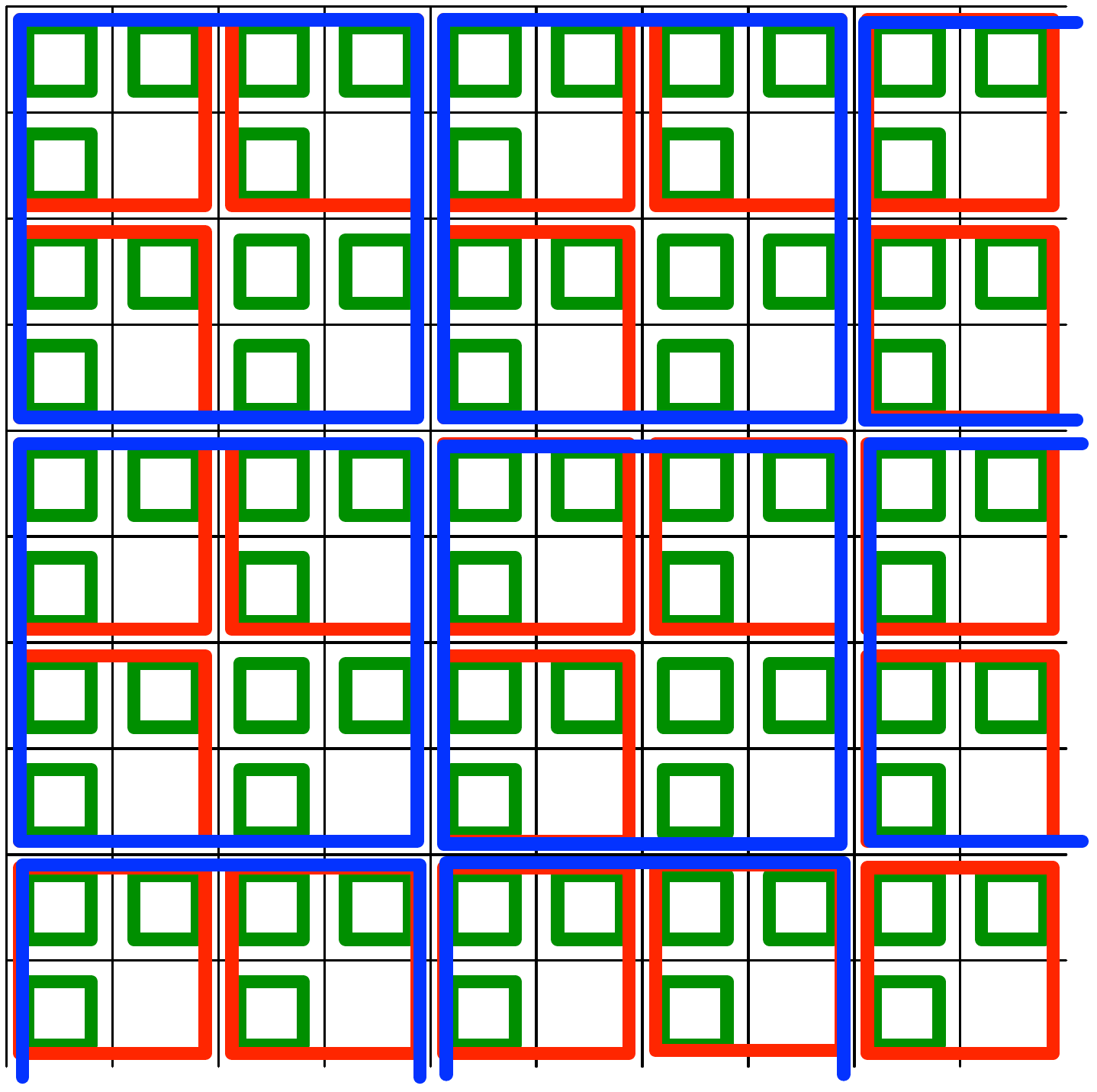}
\caption{Structure of the RG cells. A unit cell is composed of four regions (unit cells of the previous RG iteration). In each unit cell (red square), the charge of only three of the four regions is measured (green squares); the south-east corner is not measured,  leaving the total charge of the unit cell undetermined. This missing measurement is replaced at the following RG iteration by a measurement of the entire unit cell (red square), which is now a region of a renormalized unit cell (blue square). Note that this modification of the charge measurement does not need to be implemented physically, it only reflects a change in bookkeeping.}
\label{fig:ScaleInvar}
\end{figure}

In \cite{DP09a,DP10a}, we proposed a renormalization group technique to tackle this problem. First, we break the lattice into $2\times 2$ sublattices, or ``unit cells'', as illustrated on \fig{RGAL}. Given a microscopic noise model, we can  compute the probability for the value of the current across each of the four walls [North, South, East, West, c.f. \fig{RGAR}]  of each cell, conditioned on the charge configuration observed inside this cell. This produces a probability distribution $\cP_\alpha(N_\alpha,E_\alpha,S_\alpha,W_\alpha)$ for each cell $\alpha$, where $N_\alpha,E_\alpha,S_\alpha,W_\alpha$ take values representing the possible currents.\footnote{To specify the mathematical structure of the current variables, we can choose a minimal set $\{a_1,a_2,\ldots a_k\}$ of $k$ ``elementary" charges that generate all other charges under fusion. Then, any charge can be written as $a_1^{\alpha_1}\times a_2^{\alpha_2}\times \ldots a_k^{\alpha_k}$, or more succinctly represented by the vector $(\alpha_1,\alpha_2,\ldots,\alpha_k) \in \mathbb Z^{h_1} \times \mathbb Z^{h_2} \times \ldots \mathbb Z^{h_k}$ where $h_j$ is the order of charge $a_j$, meaning that  $h_j$ copies of $a_j$ always fuse to the identity. Then, the current variables $N_A,\ S_A,\ E_A,$ and $W_A$ each take value in  $\mathbb Z^{h_1} \times \mathbb Z^{h_2} \times \ldots \mathbb Z^{h_k}$. In the case of the toric code for instance, there are two elementary charges, $e$ and $m$, and their order is 2 since $e\times e = m\times m = 1$.} 

Concretely, the presence of a charge, say, in the north-east corner of the unit cell would lead to the assignment of a probability $\cO(p)$ to a current through the northern or eastern walls, and a probability $\cO(p^2)$ for the southern or western walls, reflecting the fact that the first two cases  require only one error process while the second two cases require two error processes. Here, $p$ represents the probability of an error process such as particle creation, annihilation, or displacement. The big-$\cO$ hides multiplicative factors accounting for the distinct error processes resulting in the same currents, as well as higher order processes. In any case, these probabilities can be computed exactly given an underlying local noise model. 

After having computed these current probability distributions for every cell, we merge groups of four neighbouring unit cells  into  renormalized cells (c.f. \fig{ScaleInvar}) and iterate the procedure: we sum over all the bulk processes that lead to a given current across each of the four renormalized boundaries of each cell. This is done as explained above, except that the error probability $p$ is not uniform on the lattice, but is given by the current variables of the previous RG iteration. By successive iteration, (and assuming for simplicity that the lattice linear dimension is a power of 2) we arrive at a situation where the Northern and Western walls actually correspond to the gauge line representing the non-trivial cycles of the torus. Determining the current across these walls is equivalent to decoding, as explained above.  

The difficulty with the procedure we outlined above is that charge conservation imposes strong correlations between the current variables, so their exact joint probability cannot be computed efficiently. To see this, note that the  current variables are subject to two  constraints. (a) The sum of the current entering a cell must be equal to the total charge inside the region. This leads to a conservation equation $N_\alpha+S_\alpha+E_\alpha+W_\alpha = c_\alpha$ for each cell $\alpha$, where $c_\alpha$ is the total charge contained in $\alpha$, and is known from observation (error syndrome).  (b) The currents associated to juxtaposed walls of neighbouring cells must be equal and opposite, e.g. $S_\alpha = -N_{\delta}$ when $\delta$ is the cell directly to the south of cell $\alpha$, see \fig{RGAL}. This simply follows from the fact that, e.g. $S_\alpha$ and $N_\delta$ are actually associated to the same physical boundary. Constraints (a) correlate the variables of a given cell while constraint (b) correlate variables between different cells, so the distribution is globally correlated.

Thus, approximations are required to solve this problem efficiently, as we now explain. First, just as a matter of bookkeeping, each cell stores only the random variables associated to its northern and western walls, the other ones are redundant from constraint (b). This does not affect the correlated nature of the problem however since (a) becomes $N_\alpha+W_\alpha-N_\delta-W_\beta = c_\alpha$ [c.f. \fig{RGAL}], and (b) now says that e.g. $\cP_\alpha(N_\alpha,W_\alpha,N_\delta,W_\beta)$ and  $\cP_\beta(N_\beta,W_\beta,N_\epsilon,W_\gamma)$ must be the marginals of one global distribution $\cP(N_\alpha,W_\alpha,N_\delta,N_\beta,W_\beta,N_\epsilon,W_\gamma)$. To simplify the problem, we relax this condition to a ``mean-field" condition, demanding that the two distributions yield the same marginals along the wall they share, i.e.  $\cP_\alpha(W_\beta) = \cP_\beta(W_\beta)$, where the marginals are defined the usual way
\begin{align}
\cP_\alpha(W_\beta) &= \sum_{W_\alpha,N_\delta,N_\alpha} \cP_\alpha(N_\alpha,W_\alpha,N_\delta,W_\beta) \\
\cP_\beta(W_\beta) &= \sum_{N_\beta,N_\epsilon,W_\gamma} \cP_\beta(N_\beta,W_\beta,N_\epsilon,W_\gamma).
\end{align}
These mean-field conditions are enforced heuristically using belief propagation \cite{YFW05a}.

Since mean-field approximations are not reliable in strongly correlated systems, we make one more modification to the problem. Charge conservation imposes a hard constraint (a) to the current variables, which is unlikely to ever be fulfilled in a mean-field approximation. To circumvent this problem, we let the charge $c_A$ inside each cell fluctuate, i.e., we treat it as a random variable. To describe this procedure, recall that each unit cell is composed of a collection of four regions (i.e. unit cells of the previous RG iteration). Measuring the charge distribution inside the unit cell amounts to measuring the total charge in each of these regions, which clearly fixes the total charge of the unit cell.  In the modified procedure, we measure the charge of all but one of the regions, say the south-east region. As a consequence, the total charge of the unit cell is undetermined, which relaxes the constraints on the current variables as desired. This procedure is illustrated on \fig{ScaleInvar}. The charge of the unit cell is only fixed at the following RG iteration.

\subsection{Fault-tolerant decoding}

Our description of the problem so far assumes that the charge measurements are perfect. A realistic noise model would also include faulty measurements, i.e. every charge measurement has some probability of reporting the wrong charge. To alleviate this problem, measurements can be repeated in time. A different outcome between two consecutive measurements can then be caused either from an actual error having occurred in the time between the measurements---e.g. a particle has moved in this region---or by an error in one of the two measurements. 

\begin{figure}
  \subfigure[ ]{
  \includegraphics[width=3.5 cm]{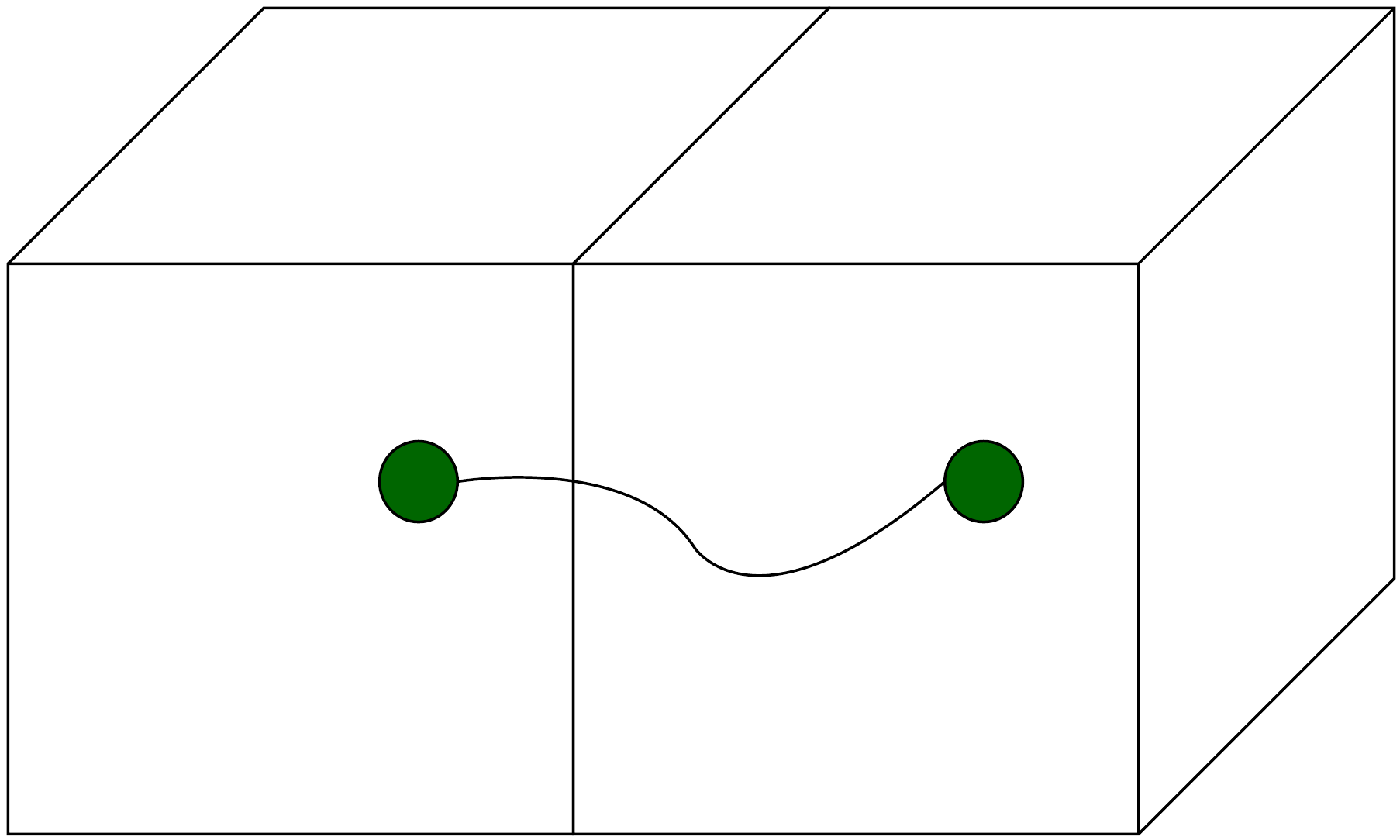}
  \label{fig:space-time-cube-a}}
  \subfigure[ ]{
  \includegraphics[width=2 cm]{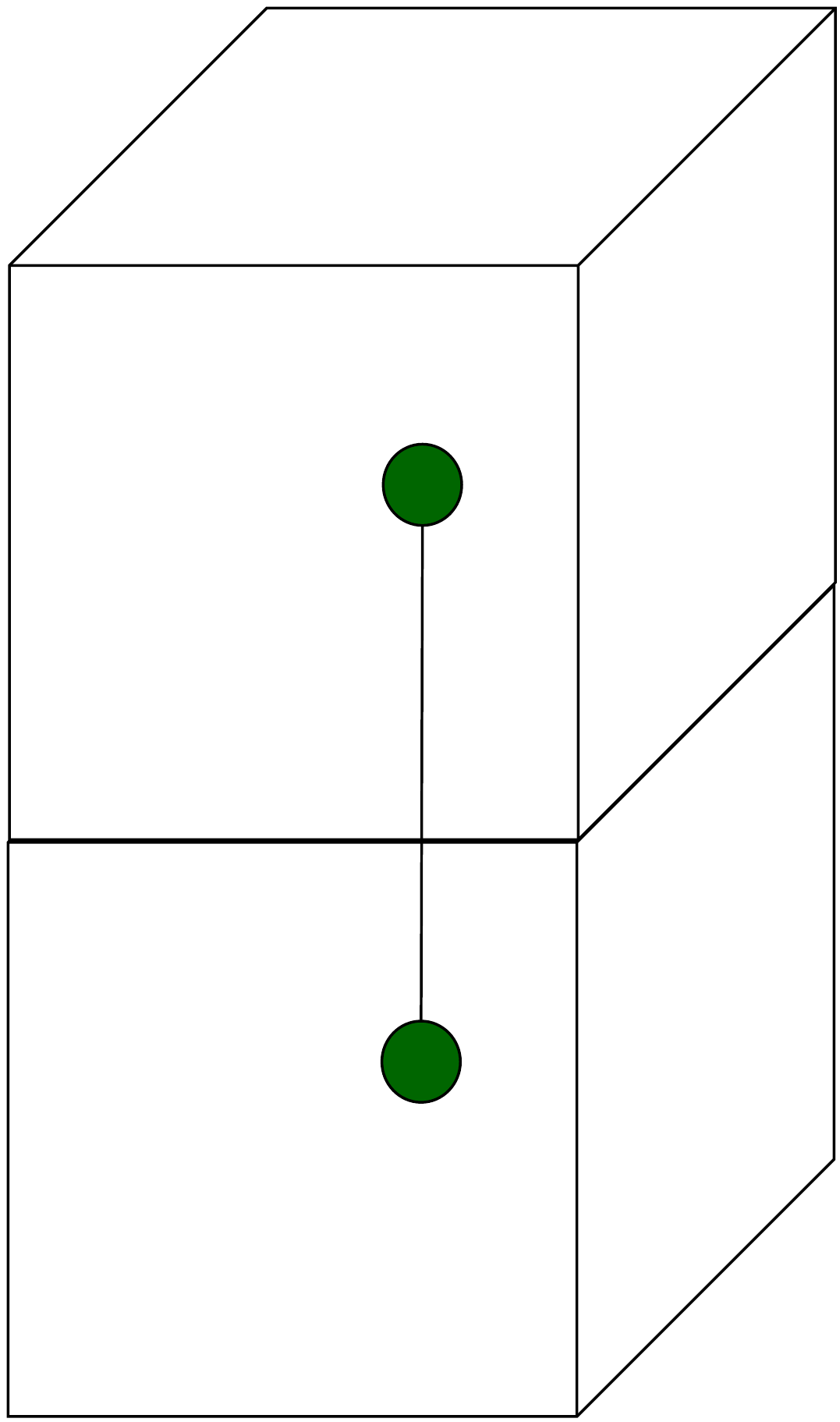}
  \label{fig:space-time-cube-b}
  }
  \caption{Space-time diagram of the fault-tolerant error-correction procedure; time flows vertically (a) A space-like error is an error that affects a qubit in between two measurements. It creates excitations in the two cubic cells with which it overlaps. (b) A time-like error is caused by a faulty measurement. It creates excitations  in the two cubic cells separated by that measurement.}
  \label{fig:space-time-cube}
\end{figure}

Consider the space-time cube enclosed between two consecutive local charge measurements (c.f. \fig{space-time-cube}). We can associate a topological charge to this cube equal to the difference between the charges revealed by the two measurements enclosing it. If the charge of a cube is non-trivial, it means that the two consecutive measurements did not yield the same result. As explained above, this could be caused by a ``space-like error" taking place between the two measurements, or a ``time-like error'' affecting the measurements themselves, see \fig{space-time-cube}. In any case, the total current across the six walls of the cube must be equal to the charge of the cube. We then see \cite{DKLP02a} that the decoding problem becomes that of determining the world-line homology of the particles in space-time.

Thus, the fault-tolerant decoding problem differs from the decoding problem with perfect measurements only in respect of the lattice dimension. Hence, the RG decoding algorithm outlined above can be applied directly. 

\section{Formal description for Kitaev's toric code}
\label{sec:KTC}

In this Section, we describe more rigorously the concepts introduced in the previous Section for the special case of Kitaev's toric code (KTC). We begin with the 2D scenario as it is technically simpler, yet conceptually equivalent to 3D. The system is a $\ell\times\ell$ square lattice, $\Lambda$, with periodic boundary conditions. We assume that $\ell$ is a integer power of 2. Each site $\Lambda_{i,j}$ $(0\leq i,j<\ell)$ holds two qubits, $\Lambda_{i,j,\alpha}$ $(\alpha\in\{H,V\}$, where $H$ and $V$ stand for horizontal and vertical, respectively). The KTC on the torus is a stabilizer code \cite{G97a} and we assume familiarity with this class of codes. 

\subsection{Model}

The stabilizer group of KTC is generated by two types of operators. On every site, $\Lambda_{i,j}$, define a \emph{site operator}, $A_{i,j}=X_{i,j,H}X_{i,j,V}X_{i,j-1,H}X_{i-1,j,V}$, and on every plaquette, define a \emph{plaquette operator}, $B_{i,j}=Z_{i,j,H}Z_{i,j+1,V}Z_{i+1,j,H}Z_{i,j,V}$ (see \fig{KTC}). Let $S_g=\{A_{i,j},B_{i,j}\}$ be the set of all plaquette and site operators. Note that it is invariant under translation. The codespace is defined to be the simultaneous +1 eigenspace of all the stabilizer operators. Equivalently, we can define the Hamiltonian $H = -\sum_{Q\in S_g}Q$, and the codespace is the degenerate ground space of $H$. There are $n = 2\ell^2$ qubits on the lattice but only $2\ell^2-2$ independent generators, i.e. $S_g$ is overcomplete. Indeed, one can easily verify that the stabilizer generators obey the two global constraints $\prod_{i,j}A_{i,j}=\II$ and $\prod_{i,j}B_{i,j}=\II$. This implies that two logical qubits are encoded in the codespace. 

The logical $X$ and $Z$ operators acting on the encoded qubits are non-trivial homological cycles (i.e loops around the torus) of $X$ operators on the dual lattice and $Z$ operators on the direct lattice. We arbitrarily choose the \emph{bare logical} operators to be
\begin{align}
	\label{eq:KTClogic}	\overline{Z}_0&=\prod_jZ_{0,j,H}&\overline{Z}_1&=\prod_iZ_{i,0,V}\\
	\nonumber		\overline{X}_0&=\prod_iX_{i,\ell-1,H}&\overline{X}_1&=\prod_jX_{\ell-1,j,V}.
\end{align}
These correspond to the gauge lines introduced in the previous section, c.f. \fig{RGAL}.

\BF
\includegraphics[width=6cm]{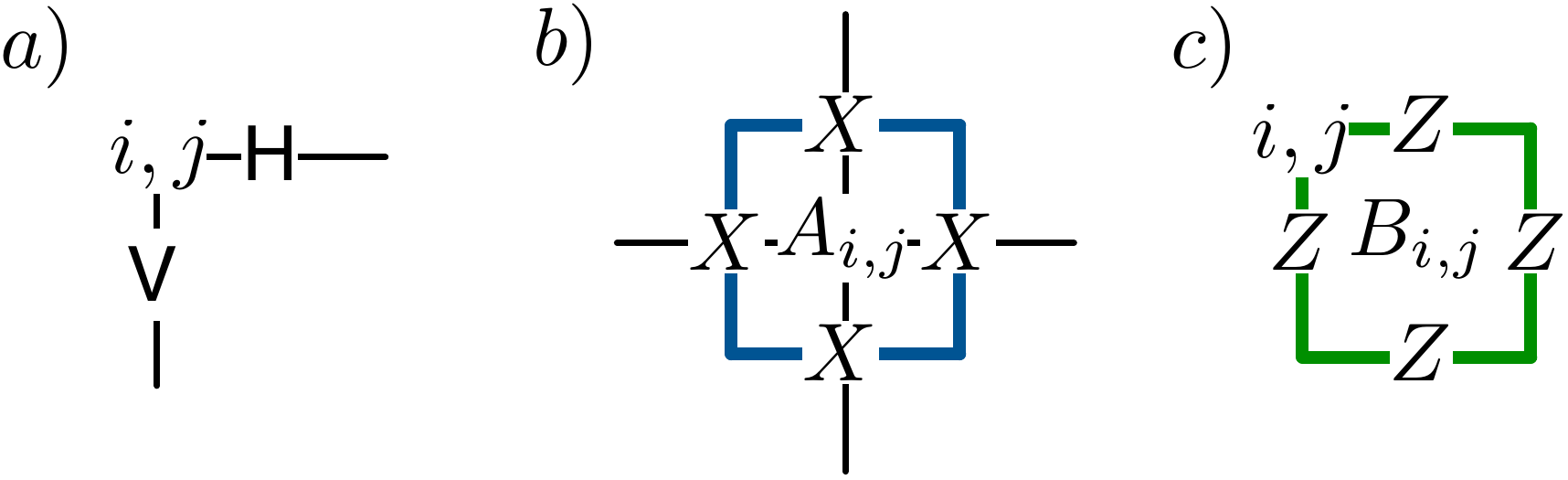}
\EF{a) One site, $\Lambda_{i,j}$, of the square lattice, $\Lambda$, on which is defined KTC. Qubits, $\Lambda_{i,j,0}$ and $\Lambda_{i,j,1}$, live on the edges and are associated to sites with the convention depicted. b) Site operator $A_{i,j}=X_{i,j,H}X_{i,j,V}X_{i,j-1,H}X_{i-1,j,V}$. Blue strings represent $X$ operators. c) Plaquette operator $B_{i,j}=Z_{i,j,H}Z_{i,j+1,V}Z_{i+1,j,H}Z_{i,j,V}$. Green strings represent $Z$ operators.}{fig:KTC}

Errors are modeled by random Pauli operators affecting the qubits. A Pauli operator will in general anti-commute with a subset of the elements of $S_g$, causing their eigenvalues to flip from +1 in the codespace to -1. An element of $S_g$ with -1 eigenvalue corresponds to a local excitation, an Abelian anyon. We refer to a plaquette excitation as a magnetic flux and to a site excitation as an electric charge. It is useful to associate binary matrices, ${\bf a}_{i,j}$ $({\bf b}_{i,j})$ to an excitation configuration, with entries 0 if the eigenvalue of $A_{i,j}$ $(B_{i,j})$ is +1 and entries 1 otherwise. Thus, the excitation configuration associated to the product of two errors is the binary sum of their respective excitation configurations---the two distinct topological charges are their own inverse. 

Since the Pauli operator $X_{i,j,H}$ anti-commutes with plaquettes $(i-1,j)$ and $(i,j)$, we see that $X$ operators can create a pair of magnetic fluxes, move a magnetic flux to a neighbouring plaquette, and annihilate a pair of neighbouring magnetic fluxes. The $Z$ Pauli operator plays an equivalent role for electric charges. Thus, the microscopic noise model describing the dynamics of the anyons can be specified by a memoryless Pauli channel $\cP_{i,j,\alpha}(Q)$, $Q\in \{I, X, Z, Y=iXZ\}$---i.e., a probability distribution over the four Pauli operators for each qubit of the lattice.  In this model, the errors $E$ affecting the system are thus elements of the $n$-qubit Pauli group $\cG^n$. The probability of an error $E = \bigotimes_{i,j,\alpha} Q_{i,j,\alpha}$ is simply given by $\cP(E) = \prod_{i,j,\alpha} \cP_{i,j,\alpha}(Q_{i,j,\alpha})$.

\subsection{Decoding problem}


When an error $E\in\cG^n$ affects the system initially in codespace, the task of error-correction is to bring the system back in the codespace by matching every excitations in pairs---thus annihilating them all---without changing the encoded information. This is realized  by applying a correction operator, $C\in\cG^n$. If the total operator $EC$ is homologically non-trivial, a logical operation will be implemented as the system is brought back to the codespace, so the information will be corrupted. To be successful, the correction $C$ must therefore be homologically equivalent to the error $E$. 

The decoding problem can be formulated in terms of this equivalence. Given an error syndrome---i.e., an excitation configuration---the decoder must find a Pauli operator that is  homologically equivalent to the error that has created this syndrome. This is a statistical inference problem. One approach to this problem is to find, among all errors that are consistent with the observed excitation configuration, the one with the highest probability. When the noise model is independent and uniform, this error is simply the lowest weight operator consistent with the excitation configuration, where the weight of $C$ is the number of non-trivial single-qubit Pauli operators in $C$. The Perfect Matching Algorithm (PMA) performs this task with a $\cO(\ell^6)$ complexity \cite{DKLP02a,H04a}.

This turns out not to be the optimal solution however. To understand this, let $t$ denote an operator with the correct excitation configuration. We suppose that $t$ is chosen in some canonical way, so it is in one-to-one correspondence with excitation configurations. The probability that the error $E$ is homologically equivalent to $t$ is simply proportional to the sum of the error probability $\cP(Q)$ over all errors $Q$ equivalent to $t$. Since the equivalence relation is generated by elements of the stabilizer group $\cS$, this is $\sum_{s\in \cS} \cP(ts)$.  On the other hand, $t$ could differ from the actual error $E$ by a combination of logical operators \eq{KTClogic}, i.e. a non-trivial cycle. Thus, we can use the group generated by the logical operators \eq{KTClogic} to label the equivalence classes of errors. Generalizing the above reasoning, the probability that the error $E$ is homologically equivalent to $tl$ defines the probability associated to the class $l\in \langle \overline X_i, \overline Z_i\rangle$ conditioned on the excitation configuration (or equivalently conditioned on $t$):
\begin{align}
	\cP(l|t) = \frac 1{\cP(t)}\sum_{s\in\cS}\cP(tls)
	\label{eq:sum} 
\end{align}
where the normalization factor is  $\cP(t) = \sum_{l,s} \cP(tls)$. The optimal decoding consists in choosing the $l$ that maximizes \eq{sum} (so the normalization $\cP(t)$ is not relevant). 
The product $tls$ is a specific Pauli operator and $\cP(tls)$ is the probability of this operator as given by the noise model. This computation is intractable because $|\cS|$ scales exponentially with the system size.  

The type of mathematical manipulation leading to \eq{sum} will be used extensively by the algorithm and in the following discussion, so Appendix \ref{sec:marginal} provides some formal background and examples that should be consulted before reading the next sections.

\subsection{RG decoding algorithm}


The RG algorithm decomposes the lattice into unit cells. We choose them to be $2\times2$ squares enclosing four plaquette and four site generators, see \fig{KTCRG}. As explained in the previous section, the RG decoder requires knowledge of all but one of the magnetic and one of the electric operators it encloses. By symmetry, we choose to leave out the south-east plaquette operator and the north-west site operator. As a consequence, the scheme will follow our description of \sec{heuristic} for the magnetic fluxes, but for the electric charges the lattice is rotated by $180^o$ relative to our description of  \sec{heuristic} . We include in the cell all the qubits that participate in the excitations measured operators, so a cell contains 12 qubits in total. Some of the qubits are shared between neighbouring cells, and this will be responsible for the constraint (b) that correlates their current variables.

\BF
\includegraphics[width=7cm]{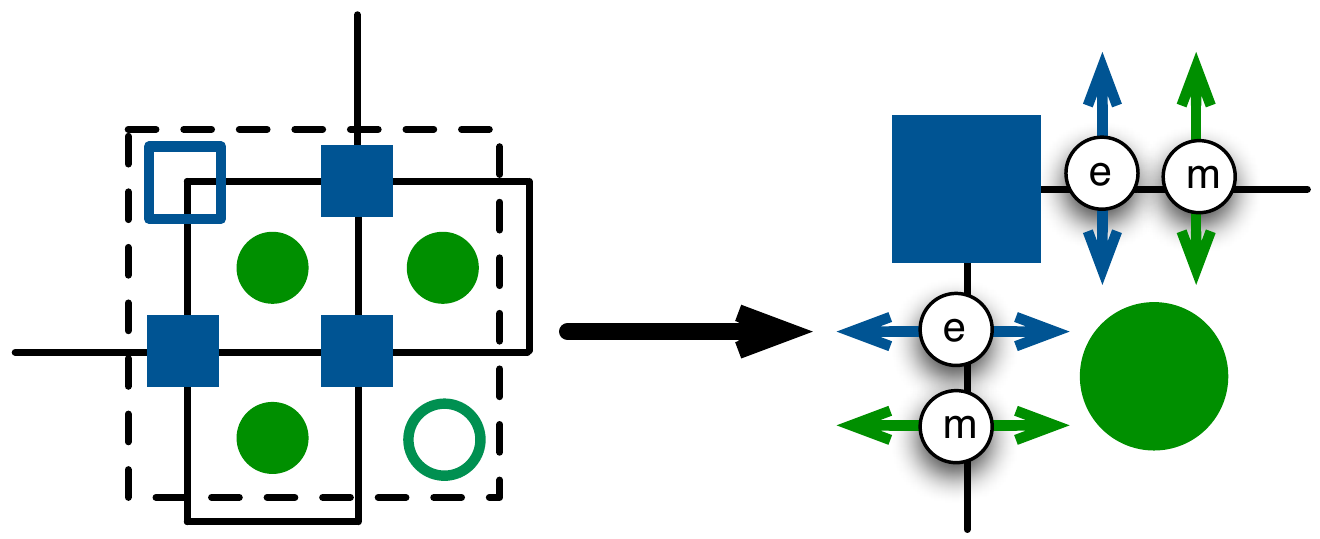}
\EF{Left: Choice of a $2\times2$ unit cell used to perfom the RG on the KTC. Green disks represent plaquette operators, blue squares represent site operators, and edges represent qubits. The two generators which are left out are represented by an empty square and a circle. Right: The RG yields a renormalized lattice. The eigenvalue of the renormalized generators corresponds to the total charge of the region and the renormalized noise model corresponds to the net flow of charges throught the boundaries (eq. \ref{eq:RG} ).}{fig:KTCRG}




To set up calculations, we define the following basis for the 12 qubits of the unit cell, see \fig{KTCUC} for qubit labeling
{\small
\begin{align}
	\nonumber		S_0&=X_0X_2X_3X_8&T_0&=Z_0&E_0&=X_6X_{10}&\overline{X}_0&=X_2X_6\\
	\nonumber		S_1&=X_1X_4X_5X_9&T_1&=Z_1&E_1&=X_7X_{11}&\overline{X}_1&=X_5X_7\\
	\nonumber		S_2&=X_3X_4X_6X_7&T_2&=Z_0Z_3&E_2&=Z_0Z_8&\overline{Z}_0&=Z_0Z_2\\
	\nonumber		S_3&=Z_0Z_1Z_3Z_4&T_3&=X_4X_7&E_3&=Z_1Z_9&\overline{Z}_1&=Z_1Z_5\\
	\nonumber		S_4&=Z_2Z_3Z_6Z_{10}&T_4&=X_6&E_4&=X_8&&\\
	\label{eq:KTCUC}	S_5&=Z_4Z_5Z_7Z_{11}&T_5&=X_7&E_5&=X_9&&\\
	\nonumber		&&&&E_6&=Z_{10}&&\\
	\nonumber		&&&&E_7&=Z_{11}&&
\end{align}}
The physical interpretations of these operators are the following. The stabilizer generators $S_j$ are the six excitation measurement operators used in the unit cell; they are plaquette and site operators. The $T_j$ are the associated canonical pure errors in the sense that $t=T_0^{a_{i,j+1}}T_1^{a_{i+1,j}}T_2^{a_{i+1,j+1}}T_3^{b_{i,j}}T_4^{b_{i,j+1}}T_5^{b_{i+1,j}}$ produces the excitation configuration ${\bf a}_{i,j}$ ${\bf b}_{i,j}$ inside the cell, without inducing any magnetic flow through the northern or western wall or any electric flow through the southern or eastern wall. The logical operators $\overline X_i$ and $\overline Z_i$ monitor respectively the magnetic current across the north ($i=0$) and west ($i=1$) wall and the electric current across the east ($i=0$) and south $(i=1$) wall. Thus, they correspond to the current variables used in \sec{heuristic}. Lastly, the $E_j$ are errors that change the charge of the site and plaquette operators that have been left out of the cell. For instance, $E_0$ brings a magnetic flux through the eastern wall into the south-east corner.
\BF
\includegraphics[width=6cm]{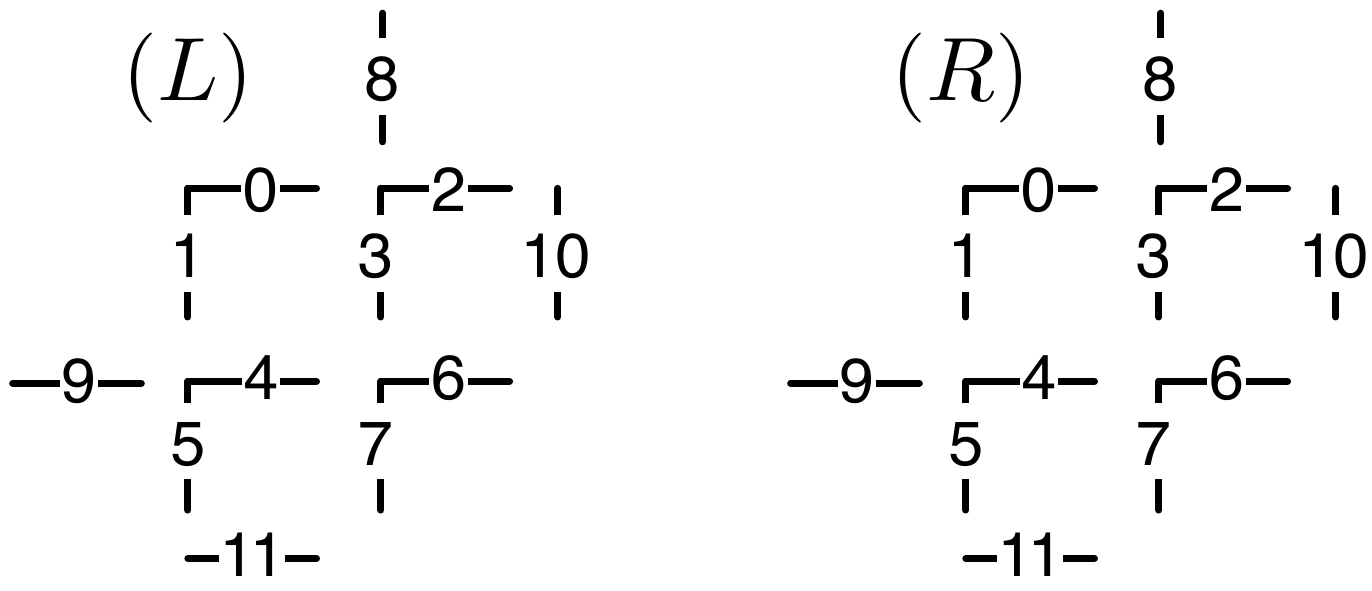}
\EF{Two neighbouring unit cells labeled $L$ and $R$. Each shows the labeling of the qubits used in \eq{KTCUC}. Note that since these two cells are neighbours, they share qubits. In particular, qubits 6 and 10 in cell $L$ are the same as qubits 9 and 1, respectively, in cell $R$.}{fig:KTCUC}

An RG iteration takes an excitation configuration and a probability distribution over the Pauli group of the qubits contained inside the unit cell, and outputs a current probability distribution obtained by summing over all equivalent processes that are consistent with the observed excitation configuration. For example, the operator $X_0$ (see \fig{KTCUC} for labeling) is equivalent to the operator $X_2X_3$ as it corresponds to a flow of one magnetic flux through the north  boundary and into the north-west plaquette. This is more directly seen when decomposed in the basis \eq{KTCUC}: $X_0=T_3\overline{X}_0S_0S_2E_4$ and $X_2X_3=T_3\overline{X}_0S_2$ since both decompose into the logical operator $\overline{X}_0$, which is associated to the magnetic current through the northern wall, and the pure error $T_3$ which is conjugate to $S_3$, the north-west plaquette. Thus, if a magnetic flux was indeed observed in the north-west corner, both of these errors should contribute to the probability of a magnetic flow through the northen boundary. More generally, the probability of a current $l \in \langle \overline X_i,\overline Z_i\rangle$ conditioned on a charge configuration $t=T_0^{a_{i,j+1}}T_1^{a_{i+1,j}}T_2^{a_{i+1,j+1}}T_3^{b_{i,j}}T_4^{b_{i,j+1}}T_5^{b_{i+1,j}}$ is given by
\begin{align}
	\label{eq:RG}\cP(l|t)\propto\sum_{s,e}\cP(tles)
\end{align}
where $s\in \langle S_i\rangle$ relates topologically equivalent trajectories and $e\in \langle E_i\rangle$ changes the value of the undetermined charge, and we left out the normalization factor $\cP(t)$.

\subsection{Belief propagation}
\label{sec:BP}




In the unit cell of Fig. \ref{fig:KTCUC} we see that there are eight qubits that belong to two unit cells; they are labeled $0, 1, 6, 7, 8, 9, 10$ and $11$. For instance, qubit 1 of cell $R$ is the same as qubit 10 of cell $L$ immediately to its left. As for any other qubits, knowledge of the excitation configuration affects the error probability of these qubits. For instance, suppose that the system contains only two magnetic fluxes, one in the north-east corner of cell $L$ and one in the north-west corner of cell $R$. In cell $L$, the presence of the magnetic flux should yield a high probability of $X$ error on qubits 2 and 10.  In cell $R$, the presence of the magnetic flux should yield a high probability of $X$ error on qubits 1 and 0. But since qubits 10 of cell $L$ and 1 of cell $R$ are actually the same, this charge configuration should globally result in a very peaked probability of an $X$ error on that qubit: it sits in between the two magnetic fluxes. But locally, given only knowledge of the charge configuration on a unique cell, this conclusion cannot be reached.

More generally, given a probability distribution over the Pauli group of the unit cell $\cP(tles)$, we can compute the marginal error probability $\cP_q(tles|_q)$ for each qubit $q$, obtained by taking a marginal of $\cP(tles)$ (c.f. App.~\ref{sec:marginal}).\footnote{The base error prior is independent on each qubit, in which case this marginal consists in the noise model on qubit $q$. But because the RG can create a correlated noise model inside a unit cell, we need this more sophisticated notion of marginal, see App.~\ref{sec:marginal}.}  When a qubit is shared between two cells, e.g. such as in the above example, its marginal conditional distributions obtained from different cells will typically differ. This is a manifestation of a violation of constraint (b) described in \sec{heuristic}. As explained there, the exact solution would be to demand that the conditional probability distribution assigned by each cell be consistent with one global probability distribution. Because of global correlations this problem is intractable, so we settle for a relaxed condition that the marginal probability distributions all agree. 


This condition is enforced by a belief propagation algorithm. This is a local message passing algorithm where messages are exchanged between neighbouring cells, there is one message per shared qubit. Initially, the outgoing messages at a cell $m^{\rm out}_q(p)$ are equal to $\cP_q(tles|_q)$ computed in that cell. These outgoing messages are then exchanged between neighbouring cells, and become incoming messages, e.g. a cell $L$ sends to its right neighbour $R$ the message $m^{\rm out}_1$ that becomes $m_{10}^{\rm in}$ in $R$,  and  receives from $R$ the message $m^{\rm out}_{10}$ that becomes $m^{\rm in}_1$ in $L$. Subsequent rounds of messages can be calculated using the received messages, following the prescription
\begin{align}
	m^{\rm out}_q(p)\leftarrow \sum_{l,s,e}\delta(tles|_q,p)\frac{\cP(tles)}{\cP_q(tles|_q)}\prod_{q'\neq q}m^{\rm in}_{q'}(tles|_{q'}),
	\label{eq:BP}
\end{align}
Here, $q,q'\in\{0, 1, 6, 7, 8, 9, 10,11\}$, $tles|_q$ is the restriction to qubit $q$ of the Pauli operator $tles$, and $\cP_q$ is the marginal on qubit $q$ of the noise model as above (c.f. App.~\ref{sec:marginal}). BP can be iterated a few times (e.g. three rounds) before executing a RG step. The messages are used to update the prior error probability, effectively replacing \eq{RG} by
\begin{align}
	\cP(l|t)\propto \sum_{e\in\langle E_0,E_1\rangle}\sum_{s\in\langle S_0S_1S_2\rangle} \cP(tles)\prod_qm^{\rm in}_q(tles|_q). \label{eq:RGm}
\end{align}

\subsection{Fault-tolerant decoding}


The prescription given for the 2D decoding problem can be applied relatively straightforwardly to the 3D problem arising from fault-tolerant decoding in the presence of faulty syndromes. To simplify the description, we will assume that there are only bit-flip errors ($X$ errors), so we only need to consider magnetic fluxes. The exact same method applies to $Z$ errors and electric charges, and moreover both type of errors can be considered simultaneously (including $Y$ errors).  

We label by $0\leq k<\tau$ the time at which the charge measurements are performed, where $\tau$ is the total duration of the computation (e.g. here we typically set $\tau = \ell$ to obtain a space-time cube). Errors affect the qubits in between measurements, and we use the label $k$ for an event that occurs in between measurement $k-1$ and $k$. There are now two types of errors to be considered. Space-like errors $\boldsymbol\eta^k$ ($\eta^k_{i,j,\alpha}\in\mathbb{Z}_2$) result in the application of the Pauli operator $E^k=\prod_{i,j,\alpha}X^{\eta^k_{i,j,\alpha}}$ to the qubits between measurements $k-1$ and $k$.  Time-like errors  $\boldsymbol\mu^k$ ($\mu_{i,j}^k\in \mathbb{Z}_2)$ result in inverting the measurement outcome at space-time coordinate $(i,j,k)$ when $\mu_{i,j}^k =1$. 


The excitation configuration measured at time $k$ results from the accumulation of space-like errors at times prior or equal to $k$, plus the measurement errors at that time, i.e. ${\bf b}^k= \boldsymbol\mu^k+\mathrm{conf}(\prod_{k'\leq k} E^{k'}) = \boldsymbol\mu^k+\sum_{k'\leq k}\mathrm{conf}(E^{k'})$. Thus, the difference between two consecutive rounds of measurements is $\Delta{\bf b}^k\equiv{\bf b}^{k-1}+{\bf b}^{k}= \boldsymbol\mu^{k-1}+ \boldsymbol\mu^{k}+\mathrm{conf}(E^{k})$. In other words, $\Delta b_{i,j}^k= \mu_{i,j}^{k-1}+\mu_{i,j}^{k}+\eta^k_{i,j,H}+\eta^k_{i,j,V}+\eta^k_{i+1,j,H}+\eta^k_{i,j+1,V}$. This defines a local space-time cubic check operator.



In this 3D picture, a $\Delta b^t_{i,j}=1$ plays the role of a magnetic flux. Note that each single error---either spatial or temporal---creates a pair of fluxes. In particular, the set of all errors can be viewed as a product of strings with magnetic fluxes located at their endpoints. 

To formalize this description, define a 3D cubic lattice of bits,  $\Lambda$, with sites $\Lambda_{i,j,k}$, holding three bits, $\Lambda_{i,j,k,\alpha}$ $(\alpha\in\{H,V,T\})$ with the convention that bits live on faces (see \fig{3DLatt}). 
The \emph{error history}, $E$, on the 3D lattice is the combination of all space-like errors $ \boldsymbol\eta$ and time-like errors $\boldsymbol\mu$, i.e. $E_{i,j,k,\alpha}=\eta^k_{i,j,\alpha}$ $(\alpha\in\{H,V\})$ and $E_{i,j,k,T}=\mu^k_{i,j}$. The excitation configuration associated to $E$ is $\Delta b^k_{i,j}$. In the following, we consider periodic boundary conditions in the spatial dimension to simplify the presentation. Then, as in the 2D case, two error histories are equivalent if they have the same excitation configuration and their product is homologically trivial on the three-torus.

The decoding problem thus stays qualitatively the same: find the most likely equivalence class of error histories consistent with the error syndrome. One subtle difference has to do with homologically non-trivial time-like loops, which do not carry the same physical meaning as space-like homologically non-trivial loops (logical operations). This difference is only caused by the unphysical boundary conditions that were chosen to simplify the presentation and the numerical simulations, and would not occur with open boundaries. In any case, a time-like logical error should not be regarded as a true memory corruption.

As in 2D, perfect matching \cite{H04a} can be used to solve an approximate version of this problem, that consists of finding, among all error histories consistent with the excitation configuration, the one with highest probability.

The optimal solution however consists in finding the most likely equivalent class of errors, and this problem can be approximated with RG techniques. The RG decoding has the same logical structure as in 2D. The lattice is broken into  $2\times 2\times 2$ unit cells. Each of these unit cells contain eight check operators (one of which is left undetermined) and 33 qubits, nine of which are shared. The current distribution over the three walls $H$, $V$, and $T$ are computed by summing over the bulk configurations consistent with a given current and excitation configuration. 
 
\BF
\includegraphics[width=4cm]{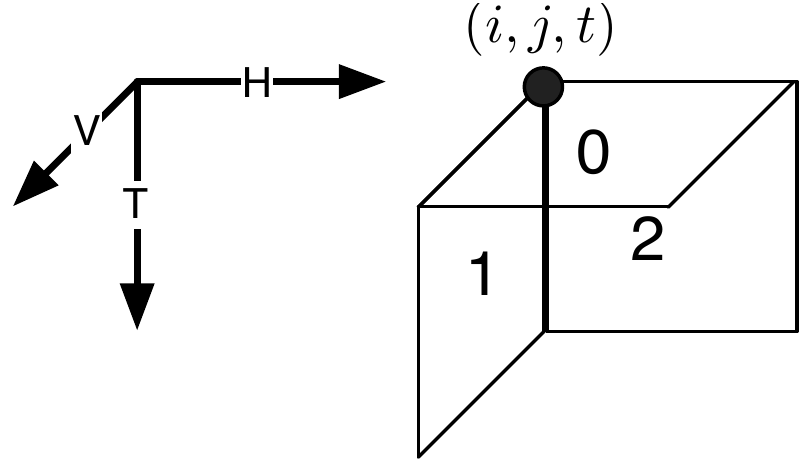}
\EF{Convention chosen for axis and unit cell of the 3D cubic lattice $\Lambda$. Bits are located on faces.}{fig:3DLatt}


There is obviously a computational cost associated to summing over the bulk processes of a unit cell. This cost grows exponentially with the number of qubits contained inside the cell. For this reason, decoding a $2\times2\times2$ unit cell involves summing over 26-bit configurations (the cell contains 33 qubits and has seven check constraints), which is fairly demanding. For this reason, we choose to work with smaller unit cells.

To make the renormalization method for fault-tolerance practical, we consider asymmetric decoding. The simplest unit cell has dimensions $2\times1\times1$ (see \fig{UC211}). In this case, the cell contains one magnetic flux operator and renormalizes only one dimension of the lattice: $\ell\times\ell\times\ell\rightarrow\ell/2\times\ell\times\ell$. For the next step, rotate the cell to renormalize another direction, e.g. $\ell/2\times\ell\times\ell\rightarrow\ell/2\times\ell/2\times\ell$. Finally, considers a second rotation to renormalize the third direction: $\ell/2\times\ell/2\times\ell\rightarrow\ell/2\times\ell/2\times\ell/2$. For this cell, we choose the following operator basis
{\small
\begin{align}
	\nonumber		S_0&=X_1X_3X_4&T_0&=X_3&E_0&=X_3X_6&L_0&=X_0X_3\\
	\label{eq:UC211}	S_1&=X_2X_3X_5&&&E_1&=X_3X_7&L_1&=X_4\\
	\nonumber		&&&&&&L_2&=X_5,
\end{align}}
with the same physical interpretation as in the 2D case.
We have also considered a $2\times2\times1$ unit cell with the following operator basis (see \fig{UC221}): 
{\small
\begin{align}
	S_0&=X_0X_3X_5&T_0&=X_5X_9&E_0&=X_9X_{12}&L_0&=X_3X_9\nonumber\\
	S_1&=X_5X_6X_9X_{11}&T_1&=X_9&E_1&=X_{11}X_{13}&L_1&=X_{10}\nonumber\\
	S_2&=X_2X_6X_8&T_2&=X_{11}&E_2&=X_5X_9X_{14}&L_2&=X_8X_{11}\nonumber\\
	\label{eq:UC221} S_3&=X_1X_4X_5&&&E_3&=X_9X_{15}\\
	S_4&=X_7X_{10}X_{11}&&&E_4&=X_{11}X_{16}\nonumber\\
	S_5&=X_1X_6X_7.\nonumber
\end{align}
}

\BF
\includegraphics[width=4cm]{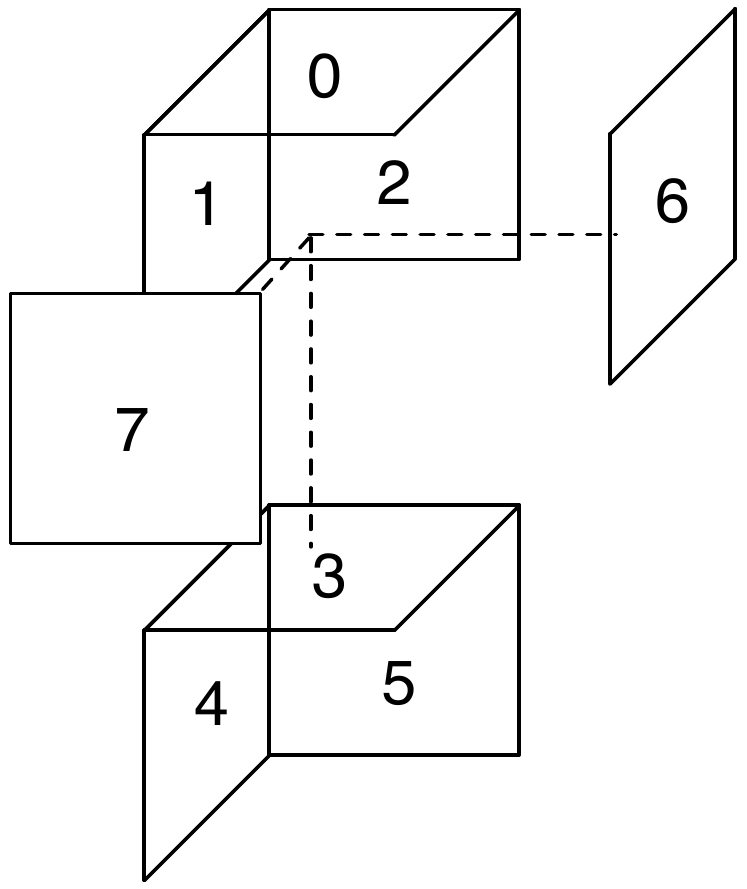}
\EF{Exploded view of $2\times1\times1$ unit cell. Qubits live on the faces. Qubits 1, 2, 6 and 7 are shared between cells and  so participate in BP. See \eq{UC211} for the operator basis.}{fig:UC211}

\section{Numerical results} 
\label{sec:numerics}

The 2D version of this RG decoding algorithm was numerically benchmarked in \cite{DP09a, DP10a} for Kitaev's toric code, and in \cite{DP13a} for the $\mathbb Z_d$ generalization of the toric code.  Here, we present numerical results obtained for the 3D fault-tolerant case (see also \cite{BDPS12a}). We consider the isotropic case where every qubit is independently subject to a bit-flip noise with probability $p$ and likewise measurements are subject to independent noise that flips their outcome with probability $p$. We use standard Monte Carlo techniques to estimate the fault-tolerant storage threshold. Our results are shown in \fig{Sim211} for the $2\times1\times 1$ cell and and \fig{Sim221} for the $2\times 2\times 1$ cell. Thresholds are observed at $p_{\rm th} \sim1.8(2)\%$ and $p_{\rm th} \sim1.9(4)\%$ respectively: for $p\leq p_{\rm th}$, the failure probability of the decoding algorithm decreases as the lattice size increases. These values should be compared to  the $2.9\%$ value obtained via PMA \cite{H04a} with the same error model.

\BF
\includegraphics[width=6cm]{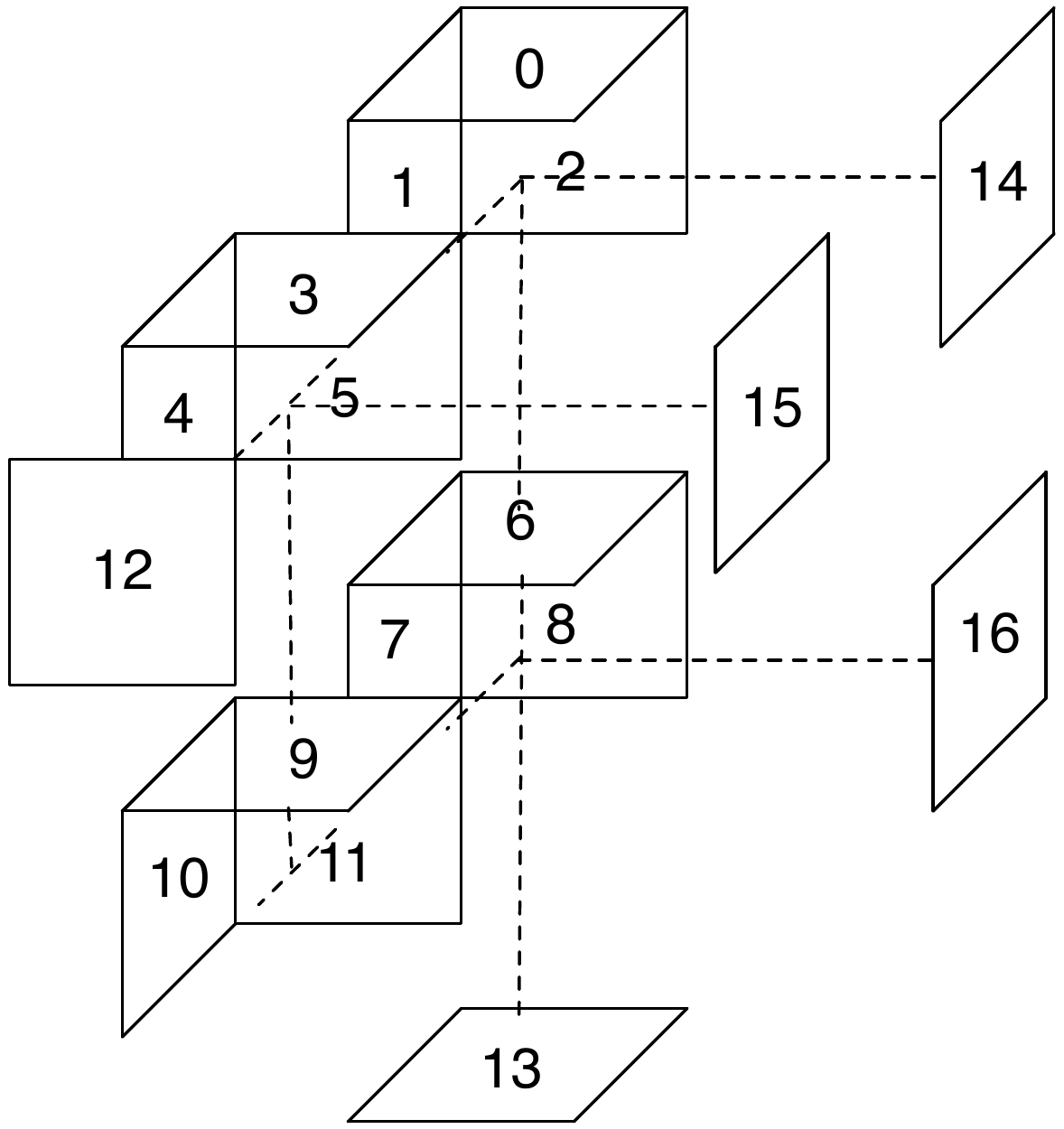}
\EF{Exploded view of a $2\times2\times1$ unit cell. Qubits 0, 1, 2, 4, 7, 12, 13, 14, 15 and 16 are shared. See \eq{UC221} for the operator basis.}{fig:UC221}

\BF
\includegraphics[width=6cm]{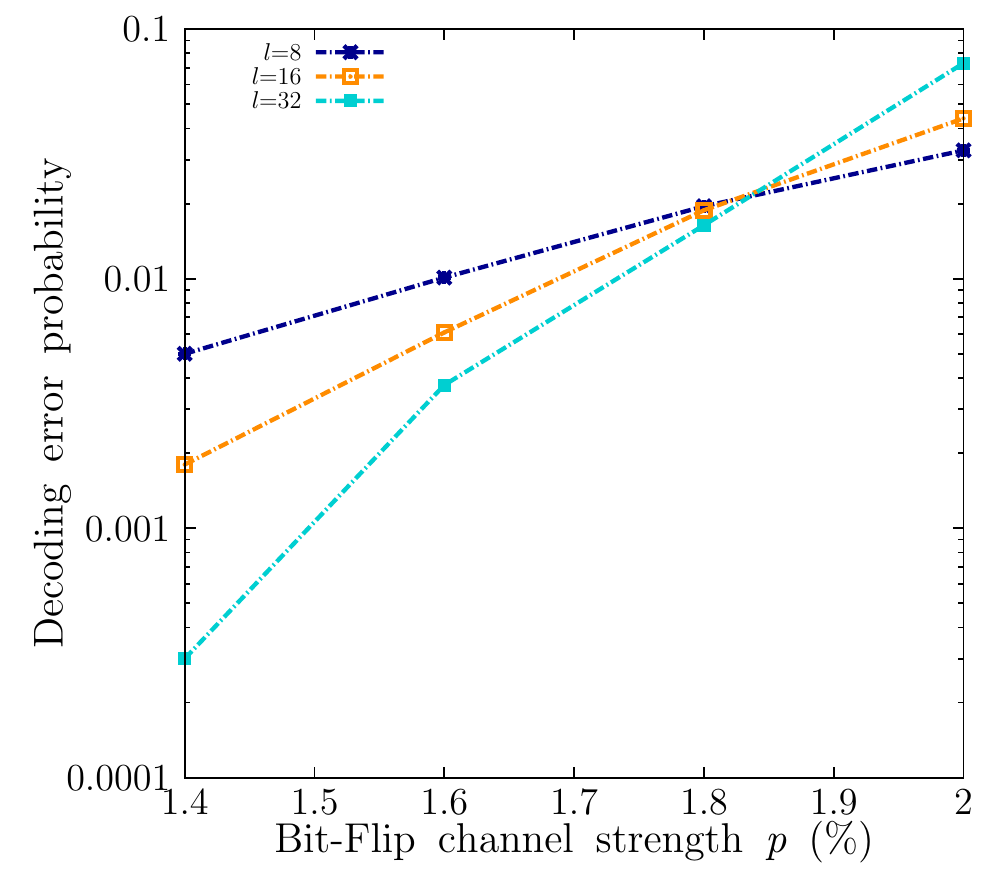}
\EF{Monte Carlo estimation of the decoding error probability as a function of bit-flip channel strength, $p$ using a $2\times1\times1$ unit cell. A threshold is observed at $\sim1.82\%$  (sample size: $10^4$ per point).}{fig:Sim211}

Note that the $2\times2\times1$ unit cell is only compatible with lattice sizes that are powers of four. Moreover, due to the large size of the unit cell,  decoding is relatively slow in this case, which limits us to small lattices $\ell=16$ and $\ell=64$  in practice.\footnote{The complexity of the RG scheme is proportional to the space-time volume of the lattice, while the complexity of PMA scales with the cube of this volume. However, the constant factor of the RG scheme is exponential with the volume of each unit cell. Although this is independent of the lattice size, the constant can be quite prohibitive for large unit cells. Note also that RG can be straightforwardly parallelized to run in time logarithmic with the space-time volume of the lattice, but we have not implemented this parallel version.}  The crossing point of the corresponding two curves gives us little confidence that we have correctly identified the threshold. For this reason, we also simulated lattice sizes $\ell=8$ and $\ell=32$ using an hybrid techniques where the $2\times2\times1$ cell was used until the very last step, where a $2\times1\times1$ call was used. The crossing point of all four curves agrees very well. This is not surprising since below threshold, we expect the error model to flow to a noiseless fixed-point, and therefore the failure rate should be largely independent of how decoding is performed at the last few RG iterations---the first RG iterations are the critical ones in determining the threshold. This observation also gives us confidence that RG could handle various lattices shapes by combining different unit cell shapes in the appropriate way deep in the RG flow.

One might suspect the threshold to be anisotropic---given the asymmetry in the RG, e.g. the direction that is renormalized first might exhibit a lower threshold. We analyzed the data by looking at the marginal error rate in the three different directions and saw no significant anisotropy. It is possible that the threshold is insensitive to such details, but that they have more subtle effect such as leading to different scaling exponents. In both cases, better statistics would be needed to give a quantitative answer.

\BF
\includegraphics[width=6cm]{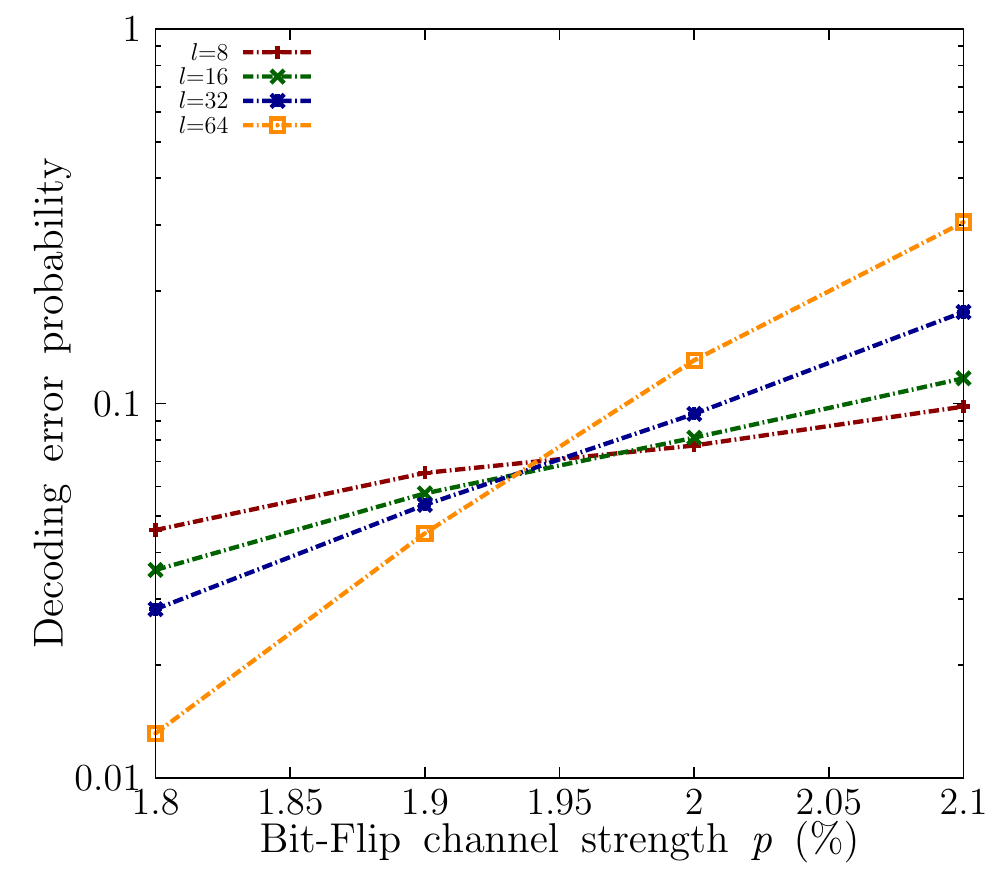}
\EF{Monte Carlo estimation of the decoding error probability as a function of bit-flip channel strength, $p$ using a $2\times2\times1$ unit cell. A threshold is observed at $\sim1.9(4)\%$. We note some finite size effects for $\ell=8$, but not for the other three curves. Sample size varies from $3\times10^3$ to $10^4$.}{fig:Sim221}

\section{Conclusion and outlook}
\label{sec:conclusion}

We have given a detailed presentation of a renormalization group algorithm for fault-tolerant decoding of  topological quantum error-correcting codes supporting Abelian anyonic excitations. This extends our previous work \cite{DP09a, DP10a} in an essential way, permitting error correction in the presence of faulty measurements. We have numerically benchmarked this algorithm and found that it achieves a fault-tolerant error threshold of nearly 2\%, in the same ballpark as the other leading techniques.

\subsection{Relation to other work}

Since the publication of our algorithm \cite{DP09a, DP10a}, there has been a number of decoding algorithms proposed for topological codes that we now briefly review. 

Sarvepalli and Raussendorf (SR) \cite{SR12a} have conceived a RG decoder for topological color codes that resembles ours in many ways. To our understanding their algorithm is conceptually identical to ours. Their presentation differs in one central way. Because some stabilizer generators unavoidably overlap with two different cells we were forced to share qubits between unit cells, which led to inter-cell correlations. Instead of this, SR split those stabilizer generators into two parts, each supported on a unique cell, and assign a binary random variable to the value of each part. The sum of these two random variables must equal the value of the syndrome associated to the stabilizer. These auxiliary binary variables play a role analogous to the shared qubits in our description. For the color code, their decoder achieves a threshold of 7.8\%, compared to 8.7\% achieved by mapping the code to multiple copies of KTC and decoding them with our RG algorithm  \cite{BDP11a}.

Bravyi and Haah (BH) \cite{BH11a} have proposed a RG decoder suitable for any topological code supporting localized Abelian anyons. It crucially differs from our approach by being based on {\em hard decisions}, while our approach uses {\em soft decisions}. In other words, the optimal recovery is only decided at the very last step of our RG iterations. At intermediate iterations, probabilities are assigned to various recoveries, but none of the options is ever ruled out until the very end. In contrast, in the BH scheme, decisions are taken to fuse certain pairs of excitations at intermediate iterations of the RG scheme. Hard decoders are conceptually simpler, and so lend themselves to more rigorous analysis. Indeed, BH were able to prove that their decoder achieves a finite threshold, while we can only provide numerical evidences for our algorithm. On the other hand, it is well known in classical coding theory that soft decoders achieve better performances \cite{P95a}. In the quantum setting, it has been shown that soft decoder can achieve a higher threshold and greater noise suppression below threshold \cite{P06a}. Their algorithm achieves a threshold of 6.7\%.

Wootton and Loss (WL) \cite{WL12a} used Monte Carlo sampling to estimate the sum in \eq{sum}, thus directly estimating the probability of each equivalence class of errors conditioned on the error syndrome. Since Monte Carlo is exact within statistical error, given a sufficiently large sample, this technique is optimal and consequently outperforms all other decoding algorithms. Indeed, they achieve a threshold of 18.5\%, compared to 16.4\% using our method with the same noise model. Its main drawback is that it is very slow compared to other methods, its runtime scales (morally) exponentially with the lattice size. 

Lastly, Fowler, Whiteside and Hollenberg (FWH) \cite{FWH12a} have implemented a parallelized version of Edmonds' perfect matching algorithm \cite{E65a} (PMA), which was the first algorithm used to decode topological code \cite{DKLP02a}.  This implementation runs in constant average time without any performance loss compared to the original PMA. Our understanding of this algorithm is that it is of Las Vegas type, meaning that its run-time is not pre-determined. For instance, in this parallel implementation, it is possible that one node of the cluster requires more time than other nodes. On a very large lattice, these fluctuations could become important, i.e. the probability that at least one node takes a time superior or equal to any finite $T$ approaches one. Thus, care must be taken in the interpretation of this constant average runtime. 

\subsection{Extensions}

It is possible to combine these techniques in various ways to obtain tradeoffs between runtime and error correction. For instance, the RG algorithm of BH can conceptually be seen as a degradation of our algorithm where probabilities on current variables $\cP(l)$ are rounded up to the closest binary distribution 
\begin{equation}
\cP'(l) = \left\{
\begin{array}{ll}
1 & {\rm if}\ l \ {\rm maximizes}\ \cP(l) \\
0 & {\rm otherwise}
\end{array}\right. .
\end{equation}
Because of this simplicity, it is much faster than our algorithm. There exist intermediate degradations that could interpolate between these two extreme schemes. For instance, we could round up the distribution to the closest trinary distribution
\begin{equation}
\cP'(l) = \left\{
\begin{array}{ll}
1 & {\rm if}\ \cP(l) \geq 1-\epsilon \\
0  & {\rm if}\ \cP(l) \leq \epsilon \\
F & {\rm otherwise}
\end{array}\right. 
\end{equation}
where the flag symbol $F$ is used to signal a potential error. Such a scheme was used by Knill \cite{K05a} in the context of concatenated codes, which can be seen as a degradation of the scheme of \cite{P06a} that uses the exact probability distribution. One could also consider keeping only the first few largest probabilities, and rounding all other to zero. 

As we have seen, larger unit cells lead to better error correction, but the exact summation \eq{RG} of bulk processes inside a unit cell scales exponentially with the volume of the cell. One possibility would be to sum the bulk processes inside the unit cell only approximately. This would enable RG decoding using much larger unit cells. For instance, we could use WL's Monte Carlo's scheme to estimate this sum. Alternatively, we could use tensor-network techniques \cite{H08a} to approximate this sum. Even without approximations, a transfer matrix approach could be used to decrease this complexity from exponential in the area of the cell (or volume in 3D) to exponential in its linear size (or area in 3D). For the small cells we considered here, these more elaborate techniques are of no use.

Lastly, we note that the description of our algorithm presented in \sec{heuristic} applies equally well to subsystem codes \cite{P05b} that have local stabilizer generators in 2D, such as the topological subsystem color codes \cite{B09a} (but excludes e.g. Bacon-Shor codes \cite{B06a}). Indeed, the stabilizer generators of these codes reveal excitations that carry topological charges and the decoding problem consists of inferring the world-line homology of these excitations. The main difference is that not all topological charges can corrupt the encoded information. Some of the topological charges---that we called {\em gauge charges} in \cite{BDP11a}---can be dragged along a non-trivial cycle without changing the ground state of the system. Thus, the current associated to these charges does not need to be monitored. Thus, \eq{RG} should contain an extra sum corresponding these harmless processes. We have used this technique for the topological subsystem color code in \cite{BDP11a} and obtained a threshold of 1.95\%.

\subsection{Acknowledgements}

We would like to thank Arvin Faruque  for useful discussions. Computational resources were provided by Calcul Qu\'ebec and Compute Canada. This work was partially funded by NSERC and by Intelligence Advanced Research Projects Activity (IARPA) via Department of Interior National Business Center contract D11PC20167. The U.S. Government is authorized to reproduce and distribute reprints for Governmental purposes notwithstanding any copyright annotation thereon. Disclaimer: The views and conclusions contained herein are those of the authors and should not be interpreted as necessarily representing the official policies or endorsements, either expressed or implied, of IARPA, DoI/NBC, or the U.S. Government.

\appendix

\section{Manipulating probabilities over $\cG^n$}
\label{sec:marginal}

In this Appendix we provide some mathematical background for manipulating probabilities over the $n$ qubit Pauli group $\cG^n$. This should be useful to understand the details of \sec{KTC} or to implement the RG algorithm.

 Let $\cP(E)$ be a probability distribution over the $n$-qubit Pauli group $\cG^n$, e.g. corresponding to a physical noise model. Given a generating set $\{Q_i\}$ of $\cG^n$, we can express any $E\in \cG^n$ as 
 \begin{equation}
 E = \prod_{i=1}^{2n} Q_i^{x_i}
 \label{eq:binary}
 \end{equation} 
where $x_i \in \{0,1\}$. This allows us to interpret $\cP(E)$ as a distribution over $2n$ binary variables $\cP(x_1,...,x_{2n}) = \cP(E = \prod_{i=1}^{2n} Q_i^{x_i})$. Standard Bayesian calculus can then be used to define marginal distributions, conditional distributions, etc. For instance, the marginal distribution over $x_1$, $x_2$, and $x_3$ is given by $\cP(x_1,x_2,x_3) = \sum_{x_4,...x_{2n}}P(x_1,x_2,\ldots x_{2n})$. The probability of $x_1$ and $x_2$ conditioned on $x_3$ is given by $\cP(x_1,x_2|x_3) = \cP(x_1,x_2,x_3)/\cP(x_3)$. These probabilities implicitly depend on a basis choice $\{Q_i\}$, and we can perform such manipulations for any basis of $\cG^n$.

These definitions extend straightforwardly to more variables. With the isomorphism \eq{binary}, we can relabel these probabilities $\cP(Q_1,Q_2,Q_3) = \cP(x_1,x_2,x_3)$, and so forth.  
 
We can create coarse grained variables associated to subgroups of the Pauli group. For instance, let $\cK = \langle Q_1,Q_2,Q_3 \rangle$ and $\cT = \langle Q_4,Q_5 \rangle$ be two subgroups of $\cG^n$. An element $K$ of $\cK$ can be decomposed as $K = Q_1^{x_1}Q_2^{x_2}Q_3^{x_3}$, and similarly an element $T$ of $\cT$ can be decomposed as $K = Q_4^{x_4}Q_5^{x_5}$. The joint, marginal, and conditional probabilities can then be defined in a natural way
\begin{align}
\cP(K,T) &= \cP(KT) = \cP(x_1,x_2,x_3,x_4,x_5) \\
\cP(K) &= \cP(x_1,x_2,x_3) \\
\cP(T) &= \cP(x_4,x_5) \\
\cP(K|T) &= \cP(x_1,x_2,x_3|x_4,x_5).
\end{align}
These are the formal definitions behind Eqs.~(\ref{eq:sum},\ref{eq:RG},\ref{eq:BP}).

Lastly, we can convert any of these probabilities---joint, marginal, and conditional---to a different basis. For instance, let $\langle Q_1',Q_2',Q_3'\rangle$ be a different generating set for $\cK$. We can express these generators in terms of the previous ones $Q_i' = \prod_{j=1,2,3} Q_j^{y_{ij}}$ with $y_{ij} \in \{0,1\}$. Suppose that we have computed $\cP(K|T) = \cP(x_1,x_2,x_3|x_4,x_5)$ using the basis $\{Q_i\}$, and now wish to compute $\cP(K|T)$ for $K = Q_1^{\prime z_1} Q_2^{\prime z_2} Q_3^{\prime z_3}$. Since 
\begin{align}
K &= \prod_{i=1,2,3} (\prod_{j=1,2,3} Q_j^{y_{ij}})^{z_i} \\
&= \prod_{j=1,2,3} Q_j^{\sum_{i=1,2,3} y_{ij}z_i} ,
\end{align} 
we see that $\cP(K|T) = \cP(z_1,z_2,z_3|x_4,x_5)= \cP(x_1,x_2,x_3|x_4,x_5)$ for $x_j = \sum_{i=1,2,3} y_{ij}z_i$. These probabilities can then be used to compute marginals over a subgroup of $\cK$ specified in terms of the primed generators. For instance, for $F \in \langle Q_1',Q_2'\rangle$ we have $\cP(F|T) = \cP(z_1,z_2|x_4,x_5)$. Thus, we see the usefulness of performing basis changes: it is used to adapt the probability to the particular subgroup we are interested in.

We will be using this type of manipulation in the special case where the basis $\{Q_j'\}$ actually corresponds to the basis of single qubit Pauli operators $\{X_i,Z_i\}$. In that case, for $K = \prod_i X_i^{\alpha_j} Z_i^{\beta_i}$ we will be using the special notation $K|_q$ to represent $X_q^{\alpha_q}Z_q^{\beta_q}$, i.e. the Pauli operator on qubit $q$ in $K$. These are the formal definitions behind many mathematical expressions of Subsection~\ref{sec:BP}.

\bibliography{3DRGv6.bib}

\end{document}